\documentclass{IEEEtran}

\usepackage{cite}
\usepackage{amsmath}
\usepackage{amssymb}
\usepackage{amsfonts}
\usepackage{algorithmic}
\usepackage{graphicx}
\usepackage{textcomp}
\usepackage{xcolor}

\def\ar{\begin{array}}
 \def\arr{\end{array}}
\def\be{\begin{eqnarray}}
 \def\en{\end{eqnarray}}
\def\bee{\begin{equation}}
 \def\ee{\end{equation}}

\newcommand{\ZTEf}{Z^{\text{TE}}_{1^-}}
\newcommand{\ZTEs}{Z^{\text{TE}}_{1^+}}
\newcommand{\ZTEff}{Z^{\text{TE}}_{2^-}}
\newcommand{\ZTEss}{Z^{\text{TE}}_{2^+}}
\newcommand{\Zair}{Z_{\text{air}}}
\newcommand{\Zdiel}{Z_{\text{sl}}}
\newcommand{\ZS}{Z_{\text{s}}}
\newcommand{\YS}{Y_{\text{s}}}
\newcommand{\kdiel}{k^{\text{sl}}_z}

\def\BibTeX{{\rm B\kern-.05em{\sc i\kern-.025em b}\kern-.08em
    T\kern-.1667em\lower.7ex\hbox{E}\kern-.125emX}}
\begin{document}
\title{Wide-Angle Reflection Suppression of Dielectric Slabs Using Nonlocal Metasurface Coatings}

\author{Alexander Zhuravlev, Sergei Kuznetsov, Daria Kiselkina, Amit Shaham, \IEEEmembership{Graduate Student Member, IEEE},  Ariel Epstein, \IEEEmembership{Senior Member, IEEE} and Stanislav Glybovski

\thanks{Copyright was transferred to IEEE.}

\thanks{This work was supported by the Russian Science Foundation (Project No. 25-19-00712, https://rscf.ru/en/project/25-19-00712/). The authors acknowledge the Shared Equipment Center CKP “VTAN” (ATRC) of the NSU Physics Department for instrumental support.}

\thanks{A. Zhuravlev, D. Kiselkina, and S.  Glybovski are  with School of Physics and Engineering, ITMO University, St. Petersburg, Russia (e-mails: a.zhuravlev@metalab.ifmo.ru,  daria.kiselkina@metalab.ifmo.ru, and s.glybovski@metalab.ifmo.ru).}
\thanks{S. Kuznetsov is with Institute of Automation and Electrometry SB RAS, Koptyug Ave. 1, 630090 Novosibirsk, Russia (e-mail: SAKuznetsov@nsu.ru).}
\thanks{A. Shaham and A. Epstein are with Andrew and Erna Viterbi Faculty of Electrical and Computer Engineering, Technion–Israel Institute of Technology, Haifa 3200003,
Israel (e-mails: samitsh@campus.technion.ac.il and epsteina@ee.technion.ac.il).}

}

\maketitle

\begin{abstract}
Any discontinuity of constitutive parameters along a wave propagation path causes scattering. For a plane wave incident onto a flat dielectric slab, reflection becomes strongly dependent on the incident angle as the electrical thickness becomes large. This behavior limits the applicability of conventional single- and multilayer anti-reflective coatings. Recently, inspired by the generalized Huygens' condition, synthesized admittance sheets implemented as metasurfaces with local response have been shown to remove reflection from a dielectric slab in a wide range of incident angles, applicable, however, only in the case of small optical slab thickness. In this work, we study plane wave transmission through dielectric slabs with arbitrary thickness coated from both sides by identical metasurfaces with nonlocal response, whose grid impedance is angularly dependent (spatially dispersive). Nonlocality is shown to play a key role in obtaining wide-angle reflection suppression in the case of optically thick slabs. To validate our approach, we study a metasurface realization composed of Interconnected Split-Ring Resonators that approximates the predicted spatial dispersion law. As demonstrated numerically and experimentally, such properly devised nonlocal metasurface coatings indeed provide transmittance enhancement (and reflectance suppression) of thick dielectric slabs across a broad range of angles, paving the path to optically thin and easy-to-fabricate anti-reflective coatings efficiently operating in a wide angular range even for optically thick dielectric slabs.
\end{abstract}


\section{\label{sec:INTR}Introduction}

\IEEEPARstart{D}{iscontinuity} of constitutive parameters along a wave propagation path induces scattering. In particular, strong reflection usually appears for a plane wave incident on an air-dielectric interface or a dielectric slab under grazing incident angles. This phenomenon, known as \textit{Fresnel reflection}, challenges the implementation of various radio frequency and optical devices, to counter this, non-reflective antenna radomes and lens coatings are often employed \cite{Kozakoff2009,Raut2011}.
Furthermore, undesired reflection from walls and window glasses deteriorates the propagation of electromagnetic waves in the millimeter-wave range, strongly restricting the coverage of modern telecommunications systems in urban areas
\cite{Oliveri2024}. In view of these substantial hurdles, it is no surprise that the global research community continuously seeks measures to reduce Fresnel reflection for various scenarios of electromagnetic wave propagation. One conventional way to reduce reflections from dielectric slabs is to use anti-reflective coatings (also known as impedance matching layers), i.e., bulky dielectric layers covering the boundaries of the initial slab and operating as quarter-wave transformers for incident plane waves. However, such devices can efficiently work only in a small range of incident angles, since the quarter-wave resonant frequency is directly related to the phase accumulation along the angle-dependent optical path within the stratified media.

To achieve wide-angle reflection suppression of a dielectric slab at a given frequency, several approaches have been introduced in the literature. One of the approaches requires multilayer anti-reflective coatings \cite{Rabinovitch1975,Herrmann1985,Nagendra1988,Sahouane2018,Hossain2021}, which are, however, bulky and difficult to implement, especially at radio frequencies. Alternatively, optically dense periodic structures of sub-wavelength engineered polarizable particles (meta-atoms) -- such as thin metamaterial layers with 3D periodicity or metasurfaces (MSs) with 2D periodicity -- can be used as compact (optically thin) wide-angle anti-reflective coatings (see, e.g., \cite{Dobrowolski2002,Chen2010,Cameron2015,He2018,Im2018,Shaham2024,Shaham2025}). In particular, in \cite{Im2018} a generalization of the quarter-wave anti-reflective layer called \textit{the universal impedance matching layer} has been presented. The layer possesses spatially and temporally dispersive permittivity and permeability tensors, enabling perfect transmission of white light independent of its polarization and propagation direction. However, the suggested method cannot lead to a planar (optically thin) matching layer. When the length of the matching layer goes to zero, its required constitutive parameters become infinite.

More recently, in \cite{Shaham2024} an analytical framework was proposed to realize the so-called generalized Huygens’ condition (GHC) that guarantees omnidirectional reflection suppression of synthesized structures. As a practical realization platform, a structure formed by a dielectric slab coated by two uniform admittance grids was analytically shown to satisfy the derived GHC to a good extent, suppressing reflections across a broad angular range. Each sheet was realized as a single-layer MS of periodic meander lines with inductive grid admittance, which was assumed to be local (independent of the plane wave incident angle). As a result, in the lossless case, reflectionless free-space-like propagation through the slab was almost perfectly reproduced even for a line current source with close proximity to the structure. However, this method and the associated closed-form formulation is applicable only to optically thin dielectric slabs. This constraint may be found limiting for some applications, since most commercially available dielectric substrates and radomes, as well as dielectric obstacles for wave propagation in urban areas, cannot be considered optically thin in the millimeter-wave range. Therefore, it is important to extend the method of \cite{Shaham2024} to omnidirectionally remove reflections even from thick dielectric slabs. As the optical thickness increases, the angular dependence of the phase accumulated across the slab becomes more intricate and, hence, can no longer support such approximated omnidirectional transparency via local admittance-sheet coatings. As such, the required relations between the driving electromagnetic fields and the resulting surface currents gradually shift from local to nonlocal as the slab thickens.

In this work, we address this issue by introducing two identical nonlocal MSs that cover both faces of a thick dielectric slab. Nonlocality in MSs can be explained as follows. The effect in which an induced surface current density on a single-layer admittance sheet at each point is affected by the incident field at surrounding points \cite{Simovski2018}, manifests itself in spatial dispersion (SD) of the macroscopic grid admittance (i.e., in the dependence on the incident wave vector, or, the incident angle for propagating spatial harmonics). For realistic MSs illuminated with an incident plane wave, SD may result in a sharp dependence of complex transmission and/or reflection coefficients on the angle of incidence. In some MS applications, SD was found to be a parasitic effect \cite{Simovski2005,Yepes2021,Yepes2022}, which must be avoided or taken into account during the synthesis. On the other hand, weak SD (in the form of bianisotropy) and strong SD (with a strongly nonlocal response to the fields) may facilitate specialized functionalities and exotic device properties, e.g. perfect absorption at two given incident angles \cite{Zhirihin2017}, angularly stable filters \cite{Goshen2024}, achieving subwavelength resolution \cite{Belov2005,Kaipa2012}, optical signal processing \cite{Kwon2018}, perfect anomalous refraction and reflection \cite{Asadchy2016,Epstein2016}, and emulating nonreciprocal phenomena \cite{Pfeiffer2016}.

Despite the large range of applications and exceptional benefits, systematic and versatile approaches to designing prescribed SD laws are mostly absent in the literature. In an attempt to tackle this hurdle, we have recently proposed and utilized such a methodology to engineer the radiation pattern of a linear current source that illuminates a reflective nonlocal MS \cite{Zhuravlev2024}. Here, we further develop this scheme, augmenting it so as to analytically demonstrate that omnidirectional reflection suppression of an optically thick dielectric slab can be achieved by tailoring nonlocal response of two identical admittance sheets coating the slab. Moreover, we propose a practical realization geometry and numerically optimize the corresponding microstructure of the MS unit cells, approaching the desired SD law. The angular behavior of the reflection and transmission coefficients is characterized numerically and experimentally in the millimeter-wave range, verifying that, indeed, broad-angle reflection suppression is obtained through the developed approach.

The paper is organized as follows. Section~\ref{sec:THR} presents an analytical study of plane wave transmission through dielectric slabs coated from both sides with identical spatially dispersive admittance sheets and the derived condition for their SD law providing omnidirectional reflection suppression at arbitrary slab thickness. Section~\ref{sec:Antireflection coating} shows the analytical comparison of local and nonlocal admittance sheet approaches. The latter can provide full wide-angle reflection suppression even for thick slabs. Section~\ref{sec:Numerical_study} presents a numerical comparison of two corresponding practical MS coatings operating at an arbitrary considered frequency of 58 GHz in the millimeter-wave range. Section~\ref{sec:exper_valid} presents the results of an experimental validation.

\section{\label{sec:THR} Analytical Model}

\begin{figure}
  \centering
  \includegraphics[width=\columnwidth]{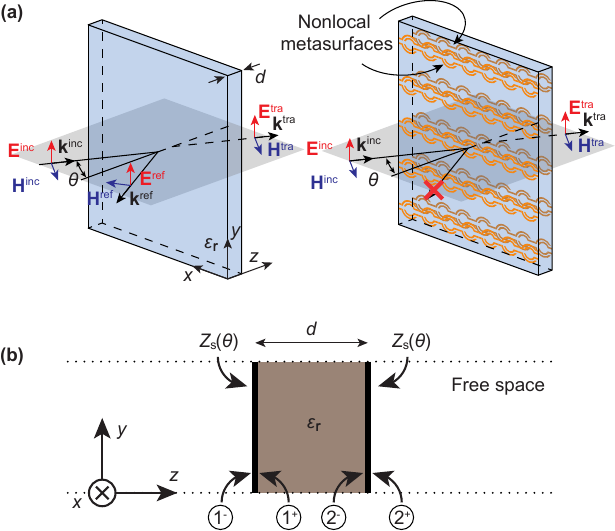} 
  \caption{Oblique incidence of a TE-polarized plane wave on a dielectric slab of thickness $d$: (a) the bare slab creating partial reflection (left) and the MS-covered slab with suppressed reflection (right) (adjacent meta-atoms of both MSs have strong capacitive coupling along the $x$ direction); (b) transfer-matrix description of the problem.}
  \label{Fig1.gen_id}
\end{figure}

To extend the method of \cite{Shaham2024}, we consider a similar boundary problem with an oblique incidence of a transverse electric (TE) plane wave ($ \mathbf{E}=E_y\cdot\mathbf{y}_0$) at angle $\theta$ on the slab which is only infinite along $x-$ and $y-$axis with relative permittivity $\varepsilon_{\text{r}}$, relative permeability of 1, and thickness $d$. Both faces of the slab are covered with identical penetrable admittance sheets, which are assumed to be realized as single-layer nonlocal MSs with strong mutual coupling between meta-atoms along the $x$ direction. The plane of incidence is the $XOZ$ plane as shown in Fig.~\ref{Fig1.gen_id}(a) for the bare slab and the MS-covered slab. For the polarization considered, the MSs possess an identical grid impedance (also referred to as sheet impedance $\ZS(\theta)=\YS^{-1}(\theta)$) which is dependent in $\theta$ \cite{Luukkonen2008}. From here forward, the symbol $\theta$ is omitted from the grid impedance designation. 

To calculate the transmission and reflection coefficients, we use the method of transfer matrices that links the tangential electric field complex amplitudes at two different reference planes $z=\text{const}$, where $z$ is the normal direction \cite[Sec.~3.4]{Collin1990}\footnote{Note that the whole structure does not create cross-polarization allowing for scalar description.}:
\begin{equation} 
\begin{pmatrix}
c_{i}\\
b_{i}
\end{pmatrix}
=
\underbrace{
\begin{pmatrix}
\frac{1}{t} & -\frac{r'}{t} \\
\frac{r}{t} & \frac{tt'-rr'}{t}
\end{pmatrix}}_{\mathbf{T}_{i,j}}
\begin{pmatrix}
c_{j}\\
b_{j}
\end{pmatrix},
\label{eq:WTM}
\end{equation}
where $c_i$ and $b_i$ ($c_j$ and $b_j$) are electric field complex amplitudes of waves traveling in positive and negative directions with respect to the $z$ axis at the $i$-th ($j$-th) reference plane. In (\ref{eq:WTM}), $t$ and $r$ ($t'$ and $r'$) denote transmission and reflection coefficients for the incident wave propagating in the positive (negative) direction. 

The transfer matrix $\mathbf{T}$ of the entire structure with a cross section schematically shown in Fig.~\ref{Fig1.gen_id}(b) can be represented as a cascade product of three matrices $\mathbf{T}=\mathbf{T}_{1^-,1^+}\cdot \mathbf{T}_{1^+,2^-}\cdot \mathbf{T}_{2^-,2^+}$. Here, $\mathbf{T}_{1^-,1^+}$ and $\mathbf{T}_{2^-,2^+}$ describe the propagation of the incident wave through a MS modeled by a sheet impedance $\ZS$ placed on a boundary between air and a dielectric half-space with the same relative permittivity as for the slab. The matrix $\mathbf{T}_{1^+,2^-}$ describes the propagation of the refracted wave within the slab \footnote{Note that in a realistic design, a fundamental mode description is valid only if $d$ is grater than the meta-atom periodicity \cite{Menzel2016}.}.

Let us first find the above matrices separately. The transfer matrix $\mathbf{T}_{1^-,1^+}$ of the first MS can be found by expressing
the following boundary conditions for an admittance sheet located at a general interface between two uniform dielectric half-spaces with relative permittivity values $\varepsilon_{r,1-}$ and $\varepsilon_{r,1+}$ \cite{Kuester2003}:
\begin{eqnarray}
\label{eq:get_matrix}
&&\mathbf{z}_0 \times \left(\mathbf{E}_{1^+} - \mathbf{E}_{1^-} \right)=0, \nonumber\\
&&\mathbf{z}_0 \times \left(\mathbf{H}_{1^+} - \mathbf{H}_{1^-} \right)=\mathbf{J}_{\text{s}}, 
\end{eqnarray}
where $\mathbf{E}_{1+,-}$ and $\mathbf{H}_{1+,-}$  are total field vectors on both sides of the sheet, while $\mathbf{z}_0$ is a unitary normal vector. The $y$-component of the induced surface current density $\mathbf{J}_{\text{s}}$ can be found as $J_y=\ZS^{-1} E_{y,1^-}=\ZS^{-1} E_{y,1^+}$ \cite{Kuester2003}.
Note that the above admittance-sheet model is valid for any single-layer MS with an electric response (formed as a patterned metal sheet with a negligible thickness) provided that the dielectric permittivity contrast between the half-spaces is low enough to neglect the substrate-induced bianisotropy and magnetic response of meta-atoms \cite{Albooyeh2015}. Using \eqref{eq:get_matrix} and applying Maxwell's equations to incident, reflected, and transmitted wave components propagating in both positive and negative directions, $\mathbf{T}_{1^-,1^+}$ can be written in the form of \eqref{eq:WTM} as:
\begin{eqnarray}
\label{eq:IMP_ON_SLAB}
&&\mathbf{T}_{1^-,1^+}= \\ 
&&\begin{pmatrix}
\frac{\ZTEf \ZTEs + \ZS(\ZTEf + \ZTEs) }{2 \ZTEs \ZS} &  
\frac{\ZTEf \ZTEs + \ZS(\ZTEs - \ZTEf) }{2 \ZTEs \ZS} \nonumber \\

\frac{-\ZTEf \ZTEs + \ZS (\ZTEs -\ZTEf)}{2 \ZTEs \ZS} 
& 
\frac{-\ZTEf \ZTEs + \ZS (\ZTEf + \ZTEs)}{2 \ZTEs \ZS}
\end{pmatrix},
\end{eqnarray}
where $\ZTEf=\eta/\sqrt{\varepsilon_{r,1^-}}\cos\theta_{1^-}$ and $\ZTEs=\eta/\sqrt{\varepsilon_{r,1^+}}\cos\theta_{1^+}$ are the wave impedances of TE polarization in the corresponding dielectric media of the half-spaces, with $\eta$ being the free space wave impedance. The angles of propagation $\theta_{1^-} \neq \theta_{1^+}$ on both sides of the MS are connected via Snell's law. The expression for the matrix $\mathbf{T}_{2^-,2^+}$ can be derived from the expression for the matrix $\mathbf{T}_{1^-,1^+}$ by the following index substitution: $1^- \rightarrow 2^-$ and $1^+ \rightarrow 2^+$. Finally, matrix $\mathbf{T}_{1^+,2^-}$ can be found as \cite[Sec.~3.4]{Collin1990}:
\begin{equation} 
\label{eq:WTM_SLAB_PR}
\mathbf{T}_{1^+,2^-}
=
\begin{pmatrix}
e^{jk_{z,1^+}d} & 0 \\
0 & e^{-jk_{z,1^+}d}
\end{pmatrix},
\end{equation}
where $k_{z,1^+}=k_0 \sqrt{\varepsilon_{1^+}}\cos\theta_{1^+}$ is the longitudinal component of the wave vector in the medium with relative permittivity $\varepsilon_{1^+}$, while $k_0$ is the free space wavenumber.  Hereinafter, the time dependence is taken in the form of $e^{j\omega t}$. For the particular case of a dielectric slab situated in air $\ZTEf=\ZTEss=\eta/\cos\theta=\Zair$, $\ZTEs=\ZTEff=\eta/\sqrt{\varepsilon_r - \sin^2\theta}=\Zdiel$, and $k_{z,1^+}=k_0 \sqrt{\varepsilon_r-\sin^2\theta}=\kdiel$.

Having calculated $\mathbf{T}$ as a matrix product, it is possible to obtain the transmission coefficient $t$ of the entire structure from the inverse of the $\mathbf{T}(1,1)$ component (see the definition of transfer matrix \eqref{eq:WTM}):
\begin{eqnarray} 
\frac{1}{t}&=\frac{j\sin(\kdiel d)\left[(\Zair\Zdiel)^2+(\Zdiel\ZS)^2+(\Zair\ZS)^2+2 \Zdiel ^2 \ZS \Zair \right]}{2\ZS^2\Zair\Zdiel} \nonumber \\
& + \frac{\cos(\kdiel d)\left[2 \ZS \Zair^2\Zdiel+2 \ZS^2\Zair \Zdiel \right]}{\ZS^2\Zair\Zdiel}.
\label{eq:trans}
\end{eqnarray}
Transmittance can be calculated from the complex transmission coefficient as $|t|^2$, while reflectance can be calculated as $|r|^2=|\mathbf{T}(2,1)/\mathbf{T}(1,1)|^2$ according to the definition of the transfer matrix \eqref{eq:WTM}.

To achieve wide-angle reflection suppression, $\ZS$ of both MSs at each incidence angle should be designed to take a suitable value such that the amplitude of $t$ equals unity, while its phase may generally take arbitrary values. For a lossless structure, this requirement guarantees that an arbitrary propagating spatial spectrum is transmitted through the slab without reflection (though with a possible wavefront shape modification). Assuming $t=e^{j\phi(\theta)}$, with $\phi(\theta)$ being an arbitrary function of the incident angle, it is possible, via substitution in \eqref{eq:trans}, to obtain a quadratic equation for the required spatially dispersive $\ZS$:
\begin{equation}
\label{eq:eqn_ZS}
\ZS^2+ \frac{\alpha_1}{\alpha_2}\ZS + \frac{\alpha_0}{\alpha_2}   = 0, 
\end{equation}
where the complex-valued coefficients read:
\begin{eqnarray}
  \alpha_0&=&je^{j\phi(\theta)}[\Zdiel \Zair]^2\sin(\kdiel d), \nonumber \\
  \alpha_1&=&2je^{j\phi(\theta)}\Zdiel^2\Zair\sin(\kdiel d) +2e^{j\phi(\theta)}\Zair^2\Zdiel \cos(\kdiel d),\nonumber \\
  \alpha_2&=&je^{j\phi(\theta)}[\Zdiel^2+\Zair^2]\sin(\kdiel d) \nonumber \\
 &+&2\Zair \Zdiel[e^{j\phi(\theta)}\cos(\kdiel d)-1].  \nonumber 
\end{eqnarray}

One can solve equation (\ref{eq:eqn_ZS}) for different desirable phase responses by setting the corresponding function of $\phi(\theta)$ to obtain the required SD law. The resulting angular dependence of $\ZS$ (the macroscopic design stage) can be then analyzed and used to select the appropriate MS realization (the microscopic design stage). 
The nomenclature regarding macroscopic and microscopic design stages in the context of MSs was introduced in \cite{Epstein2016_HMS}.

\section{\label{sec:Antireflection coating}Spatially dispersive metasurface parameters for wide-angle reflection suppression}

\begin{figure}
  \centering
  \includegraphics[width=\columnwidth]{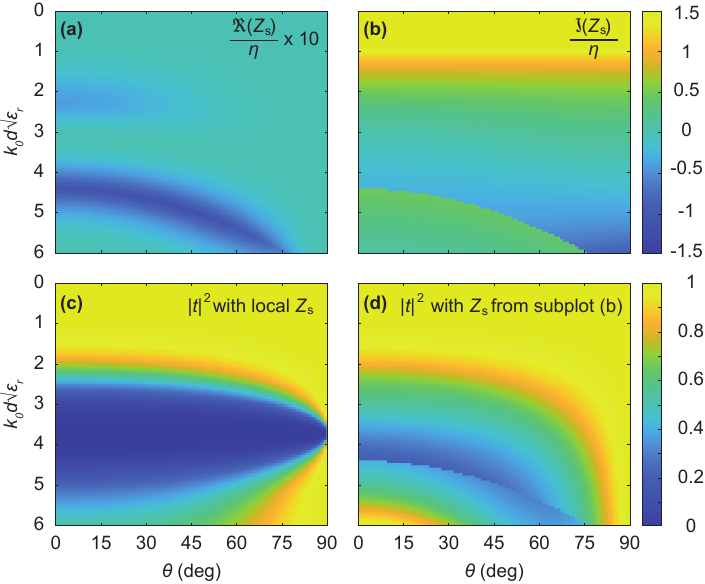} 
  \caption{Analytically calculated grid impedance values required for wide-angle reflection suppression of a dielectric slab with optical thickness $k_0d\sqrt{\varepsilon_{\text{r}}}$ and predicted transmittance levels $|t|^2$: (a) real and (b) imaginary parts of $\ZS$ for mimicking free space behavior of the slab ($t=e^{-jk_0d\cos\theta}$); (c) transmittance obtained using purely reactive and local impedance with the value of \eqref{eq:target_ZS} reached at $\theta\rightarrow 90^\circ$; (d) transmittance obtained using the imaginary part of spatially dispersive $\ZS$ according to plot (b) but omitting the real part.}
  \label{Fig2:local_nonlocal}
\end{figure}

Following the approach of \cite{Shaham2024}, let us first consider a wide-angle transparent structure that mimics free space propagation of an arbitrary wavefront, for which $\phi(\theta)=-k_0d\cos\theta$. The corresponding dependence of real and imaginary parts of $\ZS$ for different values of optical thickness $k_0d\sqrt{\varepsilon_{\text{r}}}$ parametrically calculated by solving \eqref{eq:eqn_ZS} is shown in Fig.~\ref{Fig2:local_nonlocal}(a,b). 
Note that one of two roots of \eqref{eq:eqn_ZS} attains a considerably large negative real part of $\ZS$ for all optical thicknesses (not shown); since this corresponds to a very large required gain, which is impractical, this solution has been disqualified altogether. In contrast, the other root (corresponding to the results in Fig.~\ref{Fig2:local_nonlocal}(a,b)) may have a moderate or negligible negative real part of $\ZS$ depending on thickness. It should be noted that in the limit of $\theta \to 90^\circ$, the presented $\ZS$ comes to an agreement with the solution of \cite{Shaham2024} expressed as:
\begin{equation}
    Y_\text{s}=1/\ZS=-j\eta^{-1}\sqrt{\chi_r}\tan\left({0.5k_0d\sqrt{\chi_r}}\right),
    \label{eq:target_ZS}
\end{equation}
where $\chi_r=\varepsilon_r-1$.
Fig.~\ref{Fig2:local_nonlocal}(b) shows that for small optical thicknesses of the slab (i.e., for $k_0d\sqrt{\epsilon_{\text{r}}} \ll 1$), the required grid impedance is almost constant with respect to the incident angle, taking an inductive (positive imaginary) value. In other words, no SD is required to obtain the desired reflection suppression.

The growth of the optical thickness makes the SD of $\ZS$ important to consider. Furthermore, the specific phase profile $\phi(\theta)$ chosen above demands amplification indicated by the negative real part of $\ZS$, which tends to grow with increasing thickness. 
The need for amplification is explained by the compensation of the transmitted power due to unavoidable reflection, therefore $|r|^2+|t|^2\neq1$. In other words, for arbitrary thick slabs, emulating free space with the configuration of Fig.~\ref{Fig1.gen_id}(b) becomes impossible without amplification even in the presence of SD. 

To clarify the role of SD and passivity separately, we show the calculated transmittance $|t|^2$ using \eqref{eq:trans} in two scenarios, where passivity is enforced, unlike to the case of Fig.~\ref{Fig2:local_nonlocal}(a,b). Correspondingly, in Fig.~\ref{Fig2:local_nonlocal}(c) for every optical thickness, we let $\ZS$ be constant with $\theta$ and equal to the purely imaginary value guaranteeing full reflection suppression in the grazing angle limit $\theta \rightarrow 90^{\circ}$ of \eqref{eq:target_ZS} (as proposed in \cite{Shaham2024}). This local solution can be clearly seen to operate very well for optically thin slabs, while providing poor transmission level through thick slabs at incident angles far from $90^{\circ}$. Now, let us introduce SD while still keeping the solution passive. With this aim $|t|^2$ is calculated using \eqref{eq:trans} with only the imaginary part of the spatially dispersive $\ZS$, which coincides with the values displayed in Fig.~\ref{Fig2:local_nonlocal}(b) (obtained from \eqref{eq:eqn_ZS}), and zero real part. 
The corresponding result is shown in Fig.~\ref{Fig2:local_nonlocal}(d) while some improvement with respect to non SD case is clearly seen, the overall reflection suppression is far from optimal. Therefore, omitting the required negative real part of $\ZS$ does not allow one to precisely mimic free space behavior of a coated slab with an arbitrary thickness, no matter if the MSs are local or nonlocal. Note that the resulting decrease in transmittance for both considered scenarios is especially strong in the range of incident angles and optical thicknesses, where an exact solution of \eqref{eq:eqn_ZS} predicts a small ratio $|\Im(\ZS)|/|\Re(\ZS)|$. 

To mitigate the above limitation and provide a solution using nonlocal and passive MSs, we propose to relax the requirements for the phase function $\phi(\theta)$ in \eqref{eq:eqn_ZS} while keeping the unitary transmission coefficient amplitude. With this aim, we introduce an optimization goal function as $F_{\text{goal}}=1-\overline{|t(\ZS)|}$, where the last term is the amplitude of the transmission coefficient averaged in the full range of incident angles from $0^\circ$ to $90^\circ$. At the same time, a purely imaginary value of $\ZS$ at each incident angle serves as an optimization parameter. It is worth noting that different constraints on the optimization parameters result in different realized phase functions and in different possible MS practical realizations. In our study, we use the following constraints: (i) $\Re(\ZS)=0$, to ensure passivity (that is, to avoid loss or power generation) and (ii) a reactive part for each optical thickness retains the same sign (i.e., always inductive or always capacitive) prescribed by \eqref{eq:target_ZS} for all angles of incidence. Indeed, the invariance of the sign allows one to keep the corresponding meta-atom sufficiently far from resonance at a given frequency for all incident angles. In turn, avoiding resonances in the response of the meta-atoms allows one to minimize metal and dielectric losses, which may be considerable in the millimeter-wave range. Therefore, the latter constraint simplifies the MS design at the following (microscopic) stage for a particular slab thickness. 

\begin{figure}
  \centering
  \includegraphics[width=\columnwidth]{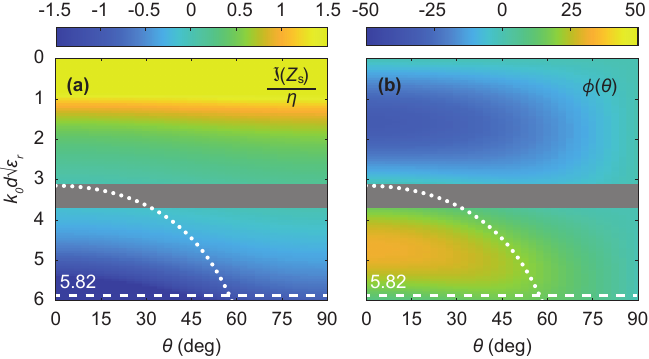} 
  \caption{Optimized macroscopic parameters of MSs coating a dielectric slab with optical thickness $k_0d\sqrt{\varepsilon_{\text{r}}}$ for wide-angle reflection suppression: (a) spatially dispersive purely imaginary grid impedance; (b) resulting phase function $\phi(\theta)$. Gray fill indicates the area in which no proper optimization solution is found with the accepted constraints. Dotted curve shows the condition of the first Fabry-Perot resonance of the bare slab. Dashed horizontal line shows the optical thickness considered for practical implementation.}
  \label{Fig3:Optimized}
\end{figure}

To calculate the corresponding purely imaginary values of $\ZS$ we used the \textit{fmincon} function of MATLAB. The initial approximation for the optimization parameters is taken from the data of Fig.~\ref{Fig2:local_nonlocal}(b). The optimization results for the purely imaginary and spatially dispersive $\ZS$, as well as for $\phi(\theta)$, corresponding to our constraints are shown in subplots (a) and (b) of Fig.~\ref{Fig3:Optimized}, respectively.

The results shown in Fig.~\ref{Fig3:Optimized}(a) and Fig.~\ref{Fig2:local_nonlocal}(b) appear to be very similar for optically thin slabs but drastically differ for the range of thicknesses and angles where neglecting the real part of an exact $\ZS$ calculated from (\ref{eq:eqn_ZS}) leads to low levels of $|t|^2$ (which can be seen in Fig.~\ref{Fig2:local_nonlocal}(c) as a blue area).
Note that for the accepted constraints, it is possible to reach the optimization goal $F_{\text{goal}}<10^{-5}$ for all optical thicknesses in the considered range of 0 to $2\pi$, except for $k_0d\sqrt{\varepsilon_{\text{r}}}\approx\pi$. This gray area in Fig.~\ref{Fig3:Optimized}(a,b) belongs to the vicinity of the first Fabry-Perot (FP) resonance of the bare dielectric slab observed from normal incidence to relatively small incident angles (see the white dotted curve). In this region, the required goal of wide-angle reflection suppression requires $\Im(\ZS)$ of the MSs to cross zero at certain angles, which contradicts the constraints taken and may lead to resonant meta-atoms with high losses. In our optimization approach, for most optical thicknesses, the MSs can remain purely inductive at all incident angles (for smaller thicknesses) or purely capacitive (for larger thicknesses).   

\section{\label{sec:Numerical_study}Numerical simulation of local and nonlocal periodic structures}

In this section, we discuss a practical implementation of the proposed spatially dispersive grid impedance for the wide-angle reflection suppression of an optically thick dielectric slab and compare its performance with the local admittance sheet approach. 

The dielectric slab to be coated has a thickness of $2.54$~mm and relative permittivity $\varepsilon_r=3.55$. These parameters are chosen to correspond to commercially available laminates compatible with standard printed circuit board (PCB) technology. The meta-atoms of the MS coatings are assumed to be etched in an 18-$\mu$m-thick perfect electric conductor sheet. The frequency of interest is $f_0=58$~GHz (in the V band used for wireless communications), which makes the slab is optically thick ($k_0d\sqrt{\epsilon_r}=5.82$ radians). This thickness is indicated with the dashed horizontal line in Fig.~\ref{Fig3:Optimized}(a), and the line corresponds to the desired SD surface impedance to be implemented. 

To solve the microscopic synthesis problem, that is, to choose the geometric parameters of meta-atoms that provide the desirable behavior of $\ZS$ at different $\theta$, we use CST Microwave Studio. The simulation model contains one meta-atom placed only on one side of the slab surrounded by Floquet periodic boundaries and illuminated with a TE-polarized plane wave with parametrically varied angle of incidence. For each practical MS coating realization, we extract $\ZS$ from the numerically calculated transmission coefficient $t$, as discussed in Appendix~\ref{appendix_A}. Next, the transmission and reflection coefficients of the slab coated from both sides with identical MS pair is analyzed using the same periodic boundaries. Note that in this section,  we omit both conductor and dielectric losses to allow proper assessment of the SD on the achievable reflection suppression of the slab.

\begin{figure*}
  \centering
  \includegraphics[width=1\textwidth]{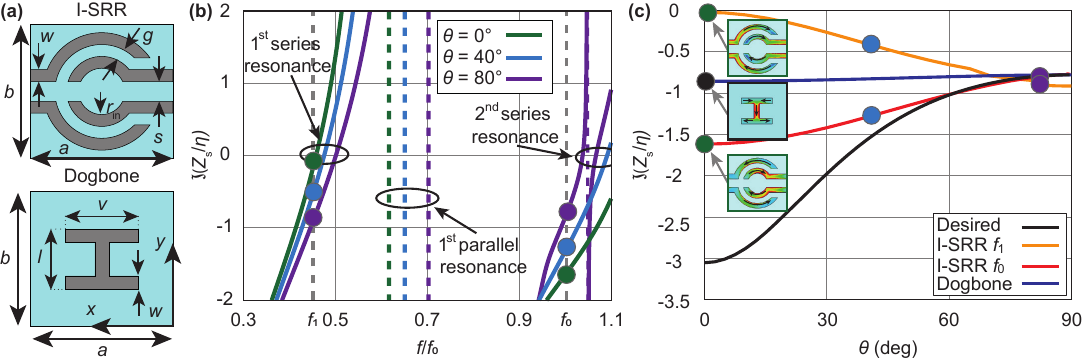} 
  \caption{Practical realization of nonlocal and local MS coatings for wide-angle reflection suppression of a dielectric slab: (a) meta-atom of the nonlocal MS coating (top) in a form of an Interconnected Split Ring Resonator (I-SRR) and a dogbone-shaped meta-atom of the local MS coating (bottom); (b) $\ZS$ vs. frequency of the nonlocal I-SRR MS coating; (c) $\ZS$ vs. incident angle of the nonlocal I-SRR MS coating in comparison with the local MS coating and the desired (optimized) curves.}
  \label{Fig4.setup}
\end{figure*}

From the prediction of Fig.~\ref{Fig3:Optimized}(a), the required spatially dispersive $\ZS$ is capacitive for all incident angles, and $|\Im(\ZS)|$ is decreasing with $\theta$.  
To enable strong SD, we consider the pattern produced by Interconnected Split Ring Resonators (I-SRRs) \cite{Ortiz} arranged with periods of $a$ and $b$ along the $x$ and $y$ directions, respectively. The meta-atoms of this coating, shown at the top of Fig.~\ref{Fig4.setup}(a), are polarized by the $y$-component of the incident electric field, producing a $y$-directed averaged surface current density.
With fixed geometric parameters $b=1.7$~mm, $g=w=0.15$~mm and $s=0.25$~mm, two other parameters ($r_{\text{in}} \in [0.25 \dots 0.35]$~mm and $a \in [1.7 \dots 1.9]$~mm) can be adjusted via numerical simulations to minimize the difference between the actual and desired curves $\Im(\ZS)$ for the entire range of incident angles. Following this approach, the optimal values found are values found are $r_{\text{in}}=0.29$~mm and $a=1.8$~mm.

The expected SD of the I-SRR MS coating is explained by strong mutual coupling between adjacent meta-atoms in the $x$ direction \cite{Ortiz}. The microscopic currents induced by the electric field along the conductors can freely flow from one SRR to another through transmission line sections. Depending on the phase difference between the currents in the inner and outer metal rings, the macroscopic capacitive grid impedance can increase or decrease with $\theta$. The frequency spectra of $\Im(\ZS)$ calculated for three different incident angles $0^{\circ}$, $40^{\circ}$, and $80^{\circ}$ are shown in Fig.~\ref{Fig4.setup}(b). It is clearly seen that in the considered range from $0.3f_0$ to $1.1f_0$ the MS coating has two series-type resonances and one parallel-type resonance. The structure exhibits an angular-dependent capacitive response at frequencies below both series-type resonances. However, in contrast to the vicinity of the first series-type resonance, where $|\Im(\ZS)|$ decreases with $\theta$, at frequencies below the second series-type resonance $\ZS$ follows the angular variation trend required in Fig.~\ref{Fig3:Optimized}(a). Hence, we prefer a frequency below the second series-type resonance, which is engineered to occur at the desired operational frequency $f_0$ as a result of parameter optimization. The realized $\Im(\ZS)$ function at $f=f_0$ is shown by the red line in Fig.~\ref{Fig4.setup}(c), while the desired curve is shown with the black one. For comparison, the angular behavior at $f=f_1\approx0.45f_0$ (below the first series-type resonance) is shown with the orange curve. The two insets for the frequencies $f_0$ and $f_1$ explain the difference in SD of the I-SRR MS coating at both series-type resonances: the currents of the inner and outer rings are induced in phase near the first resonance, while they are out of phase near the second one. It can be seen that realized $\ZS$ is close to the predicted one except for the region of small incident angles from $0^{\circ}$ to $30^{\circ}$. In fact, we could not find a way to perfectly match both curves using the available geometric parameters of the meta-atoms. However, it is more important to approximate the desired SD at large values of $\theta$ for which the reflection of the initial bare slab is stronger. 

In order to compare our results with the local admittance sheet solution of \cite{Shaham2024}, we design another capacitive structure, i.e., a dipole-type FSS with local response composed of dogbone-shaped meta-atoms (see bottom of Fig.~\ref{Fig4.setup}(a)). This coating below its first series-type resonance exhibits capacitive $\ZS$ and almost negligible SD due to weak coupling between the meta-atoms. The geometric parameters can be optimized to achieve the local impedance given by \eqref{eq:target_ZS}. With this aim, $a=b=1.5$~mm and $w=0.15$~mm are fixed, while $v$ and $l$ are numerically adjusted. The resulting behavior of $\Im(\ZS)$ corresponding to $v=0.78$~mm and $l=0.68$~mm, featuring almost negligible angular variations, is shown with the blue curve in Fig.~\ref{Fig4.setup}(c). It can be seen that the obtained curve is close to the desired black curve only at the grazing incidence. Note that in contrast to the nonlocal MS coating, the local one has a significant deviation of $\Im(\ZS)$ from the desired curve even for $45^{\circ}<\theta<80^{\circ}$. 

\begin{figure}
  \centering
  \includegraphics[width=1\columnwidth]{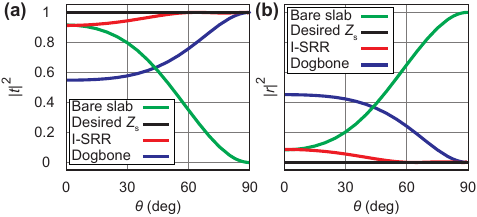} 
  \caption{Transmittance (a) and reflectance (b) of an optically thick dielectric slab as functions of the incident angle in the lossless case: black curves correspond to the slab coated with spatially dispersive admittance sheets with the desired $\ZS$; blue curves correspond to the slab coated with local MS pair with Dogbone meta-atoms; red curves correspond to nonlocal MS pair with I-SRR meta-atoms; green curves correspond to the bare slab.}
  \label{Fig5.trans_refl}
\end{figure}

The numerically calculated transmittance and reflectance as functions of $\theta$ obtained in the lossless case are shown in Fig.~\ref{Fig5.trans_refl}(a) and Fig.~\ref{Fig5.trans_refl}(b), respectively, for the bare slab (green curve), the same slab coated with the local MS pair (blue curve) and nonlocal MS pair (red curve). Similar curves for the $\ZS$ pointed in Fig.\ref{Fig4.setup}(c) are shown in black. 
It can be seen that the Dogbone MS coating provides high $|t|^2\ge0.9$
only near the grazing angles, and attains its minimal value of 0.55 at the normal incidence (the corresponding reflectance equals 0.45). 
In contrast, the proposed I-SRR MS coating provides $|t|^2\ge0.9$ ($|r|^2<0.1$) throughout the entire angular range. The residual reflection of the slab covered by I-SRR meta-atoms can be explained by the inaccuracy of the I-SRR coating impedance values in realizing the ideal desired SD of the grid impedance (compare the red and black curves in Fig.~\ref{Fig4.setup}(c)). However, a better approximation of the desired SD, e.g. by meta-atoms including additional degrees of freedom, would allow almost total transmission at all $\theta$ as predicted by the black curves in Fig.~\ref{Fig5.trans_refl}. 

\section{\label{sec:exper_valid}{Experimental validation}}

This section presents the experimental investigation of the proposed nonlocal MS pair discussed in the previous section. With this aim, the transmittance and reflectance of the MS prototype are measured as functions of the incident angle at $f=f_0=58$~GHz using a quasi-optical setup.

The prototype is manufactured as a printed circuit board composed of two laminate cores WL-CT338 [$\varepsilon_r=3.55(1-j0.004)$], with thicknesses of 1.52~mm and 0.81~mm, bonded together by two prepreg layers WL-PP350 [$\varepsilon_r=3.55(1-j0.004)$], with a thickness of 0.10~mm each. The total thickness of the dielectric slab ($d\approx2.54$~mm), its real part of the relative permittivity equals 3.55, and the dimensions of the I-SRR meta-atoms made of copper are the same as in the numerical simulations presented in Fig.~\ref{Fig5.trans_refl}. The lateral size of the MS prototype is $25\times11$~cm, which is sufficient to capture the entire incident Gaussian beam (with an aperture of $\approx30$~mm measured at a level of -10~dB from the maximum) even at oblique incidence under $|\theta|\leq 80^\circ
$.  For reference, we manufactured a prototype of a bare slab with the identical stack-up and the same size as for the MS but with no metallization. 

\begin{figure*}
  \centering
  \includegraphics[width=1\textwidth]{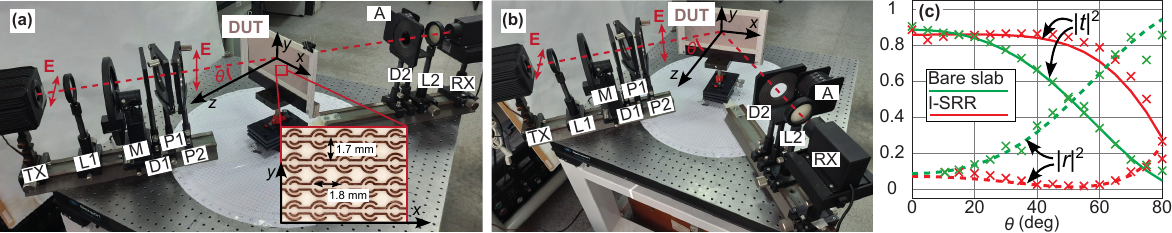} 
  \caption{To the experimental comparison of the dielectric slab coated by I-SRR MS pair with the bare slab at 58~GHz: quasi-optical schemes for measuring transmittance (a) and reflectance (b); measured transmittance  and reflectance as functions of the incident angle (markers) and the same curves calculated numerically in the presence of losses (lines).}
  \label{Fig6.exp}
\end{figure*}

For the experimental characterization of the device under test (DUT), two quasi-optical schemes are employed. The first, for measuring transmittance, is shown in Fig.~\ref{Fig6.exp}(a), while the second, for measuring reflectance, is shown in Fig.~\ref{Fig6.exp}(b).To calibrate the first scheme, the transmittance is measured in the absence of DUT in the holder, the second scheme is calibrated by measuring the reflectance of a metallic sheet. Both schemes are similar to those reported in \cite{Kuznetsov2024} and perform spectroscopy measurements in the frequency domain using a tunable backward wave oscillator (BWO) as a coherent monochromatic source \cite{Gorshunov2008}.
In this setup, the BWO (TX) generates a horizontally polarized radiation beam with a quasi-planar wavefront collimated with a phase-correcting polyethylene lens (L$_1$) mounted on its output.  After passing through the 23~Hz amplitude modulator (M), implemented as a rotating metallic disc with a diameter of 170~mm and perforation (the modulator is necessary for lock-in signal detection), the beam passes through two one-dimensional wire-grid polarizers with grids oriented sequentially at angles of $45^\circ$ (P$_1$) and $0^\circ$ (P$_2$) relative to the horizontal direction, respectively. The set of polarizers makes the beam impinging the DUT vertically polarized. The DUT is fixed on a rotating platform equipped with special screws for both coarse and fine (high-precision) angle adjustment. The platform can be manually rotated at an arbitrary angle $\theta$ about the $y$-axis, thus realizing TE-polarized wave excitation. Two absorbing diaphragms (D$_1$ and D$_2$) with a clear aperture diameter of 40 mm are placed before and after the DUT to suppress spurious side waves, which can appear upon parasitic diffraction on the quasi-optical elements of the setup and further propagate towards the detector. A Golay cell (RX) combined with a 23 Hz lock-in amplifier is used as a radiation detector, while the wave beam is focused onto the input window of the Golay cell by the second polyethylene lens (L$_2$). To avoid exceeding the saturation limit of the detector with the incoming BWO radiation, a set of thin film absorbing attenuators (A) is placed before the lens L$_2$. 
 
Fig.~\ref{Fig6.exp}(c) shows with markers the measured transmittance and reflectance of the slab coated by the I-SRR MS pair and the reference bare slab as functions of the angle of incidence. The same plot also shows with solid ($|t|^2$) and dashed ($|r|^2$) lines the same curves calculated numerically in CST Microwave Studio in the presence of metal and dielectric losses. In numerical simulations with periodic Floquet periodic boundaries, the metal traces of each I-SRR are modeled using lossy metal copper with roughness of 3~$\mu$m (a standard value for electrodeposited copper). The dielectric slab has the same relative dielectric permittivity as the materials used in fabrication. 

There is a good coincidence between the experimental results and the simulated ones, when realistic losses are taken into account. 
Some deviation of the experimental markers from the simulated curves at very large angles ($\theta>70^\circ$) can be explained by the unavoidable effects of aperture diffraction, mainly on the supplementary peripheral vertical plastic strips used to mount the DUT on the rotating platform. Comparing the I-SRR curves in Fig.~\ref{Fig6.exp}(c) and Fig.~\ref{Fig5.trans_refl}, one may notice that the impact of losses on the performance of the nonlocal prototype is considerable, reducing its effectiveness. Nevertheless, the predicted superiority with respect to the reference bare slab is maintained, especially pronounced at large angles (where Fresnel reflection issues are most severe). Importantly, the reflections caused by the coated slab remain very low across the entire angular range, highlighting that the performance deviations from the designed prototype indeed stem from unavoidable losses.


The measured data show that, despite losses, the transmittance of the slab coated with an I-SRR MS pair remains higher than $\approx0.3$ up to the incident angle of $80^\circ$ and remains higher than that of one of the bare slab almost at all angles. At the same time, the reflectance stays below $0.2$, reaching this value at $\theta=80^\circ$, corresponding to an absorptance level of $\approx0.5$. Therefore, losses in the MSs limit the desirable effect of wide-angle reflection suppression. In contrast, for the bare slab, the reflectance is lower than 0.2 only at angles smaller than $30^\circ$.

\section{Conclusion}

In this paper, the role of nonlocality in MSs coating a thick dielectric slab is studied theoretically and experimentally. The analytical formulas derived using the transmission matrix approach allow one to predict the required SD law for the macroscopic grid impedance $\ZS$ to realize a specific angular behavior of the complex plane-wave transmission coefficient for a slab of arbitrary optical thickness. 

Based on the formulas, it is shown that nonlocality that manifests itself in a pronounced angular dependence of $\ZS$ is necessary to provide wide-angle reflection suppression for large optical thicknesses of the slab (2 radians and larger). In fact, when limited to the local MS coatings, an optically thick slab can be made transparent only in a narrow range of incident angles, while wide-angle reflection suppression is only possible for relatively thin slabs \cite{Shaham2024}. 

Furthermore, it is shown that mimicking the behavior of free space in optically thick coated dielectric slabs is impossible without amplification of the transmitted plane wave, even in the presence of nonlocality in the coating MSs. However, reaching a total transmission of thick slabs with passive nonlocal MS coatings is theoretically possible by relaxing the requirement on the transmission phase. In this case, different constraints for the angular behavior of a purely imaginary $\ZS$ can be imposed, resulting in different angular variations of the transmission phase. In other words, with passive and nonlocal MS coatings one can make a thick dielectric slab wide-angle reflectionless, but at the cost of phase distortion of an arbitrary incident wave front.
As an example, with a requirement of a constant sign of $\ZS$ throughout the full angular range, the optimal SD law of a capacitive or inductive grid impedance can be obtained for thick layers, except for the vicinity of a Fabry-Perot resonance at normal incidence. In this work, for the chosen optical thickness of 5.82 radians, this approach is used to obtain a spatially dispersive and purely capacitive grid impedance, which is then used as a goal function to numerically optimize the practical realization of the MS coating. 

Two main difficulties in the practical realization of nonlocal MS coatings for wide-angle reflection suppression of an optically thick slab are: (i) power dissipation (especially losses in metal traces of meta-atoms at grazing angles) and (ii) inaccurate approximation of the  desired SD law.
The first problem can be addressed to some extent at the macroscopic design stage by setting a proper constraint to a purely reactive $\ZS$. For example, by relaxing the transmission phase, the impedance can be enforced to keep the same sign at any incident angle and to avoid small values of $|\Im(\ZS)|$ (i.e. to avoid a too close proximity of a series-type resonance). 
To address the second problem at the microscopic design stage, we propose using the MS coating composed of coupled I-SRRs. As shown in the numerical simulations, this structure allows one to approximate the desired SD only at high angles from the normal (which are, however, the most affected by the reflection in the bare dielectric slab). Despite the fact that this approximation error deteriorates the reachable reflection suppression, the performance of this nonlocal MS coating is shown to significantly exceed one of a local MS coating, either in the lossless or lossy case. Finding new practical realizations of nonlocal MS coatings with more degrees of freedom for fine SD adjustment may facilitate enhanced performance, and will be addressed in future work.

Finally, the proposed nonlocal MS pair experimentally demonstrates in the V band the possibility to extend the angular range of low reflectance (20\% and lower) from an optically thick slab from $30^\circ$ to $80^\circ$ (measured from normal). This result suggests a new useful function of nonlocal MSs, i.e. optically thin and easy-to-fabricate anti-reflective coatings efficiently operating in a wide angular range even for optically thick dielectric slabs. This function could be applied in future work to improve signal propagation in millimeter-wave telecommunication systems.

\appendices

\section{ Surface impedance extraction}
\label{appendix_A}

To extract the grid impedance $\ZS$ associated with a single-layer MS, one can numerically calculate the complex transmission coefficient $t$ through a given slab with thickness $d$ covered by the MS only from one side\footnote{We assume that nearfield coupling effects are negligible, which is naturally expected for the optically thick substrates considered herein.}.

The transfer matrix of this structure reads $\mathbf{T}^{\text{slab}}=\mathbf{T}_{1^-,1^+}\cdot \mathbf{T}_{1^+,2^-}\cdot \mathbf{T}'_{2^-,2^+}$. The first two matrices are given by \eqref{eq:IMP_ON_SLAB} and \eqref{eq:WTM_SLAB_PR}. The matrix $\mathbf{T}'_{2^-,2^+}$ describing the field discontinuity between the dielectric and air media can be obtained from \eqref{eq:IMP_ON_SLAB} by the following index substitution: $1^- \rightarrow 2^-$ and $1^+ \rightarrow 2^+$ and forcing $\ZS \rightarrow \infty$. The last condition means that the second MS on the other side of the slab vanishes. One can write: 
\begin{eqnarray}
\mathbf{T}'_{2^-,2^+}=  
\begin{pmatrix}
\frac{\ZTEff + \ZTEss }{2 \ZTEss} &  
\frac{\ZTEss - \ZTEff }{2 \ZTEss} \\

\frac{\ZTEss -\ZTEff}{2 \ZTEss} 
& 
\frac{\ZTEff + \ZTEss}{2 \ZTEss}
\end{pmatrix}.
\label{eq:IMP_JUMP}
\end{eqnarray}
Similarly to the discussion in the analytical section, using Snell’s law and the fact that some reference planes are located in the same media (see Fig.~\ref{Fig1.gen_id}(b)), it is possible to derive a simplified expression for $\mathbf{T}^{\text{slab}}$. Next, from $t=1/\mathbf{T}^{\text{slab}}(1,1)$ it is possible to find a compact relation between $\ZS$ and $t$, as follows: 
\begin{equation} 
\ZS=\frac{t e^{j\kdiel d}(\Zair \Zdiel^2+\Zair^2 \Zdiel)+t e^{-j\kdiel d} (\Zair^2 \Zdiel - \Zair \Zdiel^2)}{4\Zdiel \Zair -t e^{j\kdiel d} (\Zair + \Zdiel)^2+t e^{-j \kdiel d} (\Zdiel-\Zair)^2}.
\label{eq:app_retr}
\end{equation}

As the slab has known properties and is covered only from one side by the MS, the extraction procedure for $\ZS$ is unambiguous.



 \bibliographystyle{ieeetr}
 \bibliography{refs}

\end{document}